\documentclass[twocolumn, amssymb,nobibnotes,aps,prb,nopacs]{revtex4}

\usepackage{amsmath}
\usepackage{graphicx}
\newcommand \dd[1]  { \,\textrm d{#1}                       }   
\begin{document}

\title{Kinetic-simulation study of propagation of Langmuir-like ionic waves in dusty plasma}

\author{S. M. Hosseini Jenab\footnote{Email: Mehdi.Jenab@nwu.ac.za}}
\author{F. Spanier \footnote{Email: Felix@fspanier.de}}
\affiliation{Centre for Space Research, North-West University, Potchefstroom Campus, Private Bag X6001, Potchefstroom 2520, South Africa}

\date{\today}

\begin{abstract}

The propagation of ionic perturbations in a dusty plasma 
is considered through a three-species kinetic simulation approach, 
in which the temporal evolution of all three elements i.e. electrons,
ions and dust particles are followed based on 
the Vlasov equation coupled with the Poisson equation.
Two cases are focused upon: 
firstly a fully electron depleted dusty plasma,
i.e., a plasma consisting of ions and dust-particles.
The second case includes dusty plasmas 
with large electron-to-ion temperature ratios. 
The main features of the ionic waves in 
these two settings including the dispersion relation and the Landau damping rate are studied. 
It is shown that the dispersion relation of 
the ionic waves perfectly matches 
the dispersion relation of Langmuir waves and hence are called Langmuir-like ionic waves
and can be considered as Langmuir-like ionic waves. 
These waves can be theoretically predicted 
by the dispersion relation of the dust-ion-acoustic waves.
The transition of ionic waves from dust-ion-acoustic 
to Langmuir-like waves are shown to be sharp/smooth in first/second case.
The Landau damping rates based on simulation results
are presented and compared with theoretical predictions wherever possible.

\end{abstract}

\maketitle

\section{Introduction}
Langmuir waves have been the first kind of waves discovered in plasma physics by Langmuir \cite{Lang},
in which the electrons' oscillations propagate through the plasma media with a frequency 
$\omega_{pe} = \sqrt{\frac{n_{e0}e^2}{m_e \epsilon_0}}\ll \omega$,
is the plasma frequency for electrons.
It also possesses a dispersion relation in a restricted approximation 
of small wavenumber $k \lambda_{De}\ll 1$ (electron Deybe length 
$\lambda_{De}=\sqrt{\frac{\epsilon_0 K_B T_e}{n_{e0}e^2}}$) 
known as the Bohm-Gross equation:
\begin{equation}
\omega^2= \omega_{pe}^2 + 3\frac{K_B T_e}{m_e} k^2,
\end{equation}
in which $K_B$, $T_e$ and $m_e$ are Boltzman constant, 
the electron temperature and its mass respectively.
However more precise dispersion relations were presented over 
the years with less restriction on $k \lambda_{De}\ll 1$ 
and here the empirical equation suggested by Swanson\cite{Swason}
(see Eq.\ref{Swanson}) is considered as well. 
The dispersion relation is achieved by considering the fact 
that ions due to their large mass react too slowly 
to fast oscillations of the electric field and thus 
can be regarded as stationary and immobile.
Ions play the important role of a positive background
which maintains the local quasineutrality condition throughout the plasma medium.
Electrons on the other hand 
are responding to the electric field oscillations
providing the inertia necessary for wave propagation by their masses.
So the mechanism of Langmuir waves consists of resonating charged particles,
neutralized by oppositely charged particles in the background,
oscillating and resonating to electric field oscillation 
while the background charged particles stay passive 
to the electric field oscillation -in case of an ion-electron plasma,
due to large mass ratio of background particles to resonating particles.
Such a mechanism introduces a clear cut in dispersion relation 
called the cut-off frequency, waves with the frequency 
below the plasma frequency of the resonating charged particles 
can't prorogate in the plasma media, e.g. 
in an ion-electron plasma, waves with angular frequency $\omega < \omega_{pe}$ are damped heavily.

As for ionic waves, the fundamental waves include ion acoustic waves (IAWs)\cite{Tonks}
which consist of electrostatic waves propagating with the frequency 
$\omega_{pi}\ll\omega\ll\omega_{pe}$
in which, contrary to the Langmuir wave mechanism, both 
electrons and ions respond to the electric field oscillation.
This kind of wave can be found in dusty plasmas as well
i.e. dust-ion-acoustic waves (DIAWs) 
$kv_{Td} \ll kv_{Ti}\ll \omega \ll kv_{Te}$ 
although with a modified version of 
the ion-acoustic dispersion relation of an ion-electron plasma\cite{Bellan}:
\begin{equation}
  \omega^2=k^2 \big( \frac{n_{i0}}{n_{e0}} \frac{k_B T_e}{m_i} \frac{1}{1+ k^2 \lambda_{De}^2} + 3 \frac{k_B T_i}{m_i}\big).
\end{equation}

Dusty plasmas as an active frontier of the 
plasma research\cite{Shukla02, Vlad05, Verheest, Fortov} 
consist of three constitutes including electrons, ions and dust particles. 
The dust particles include heavy particles appearing in different plasmas
ranging from laboratory to astrophysics and coming from widely different
sources such as scrap-off particles of the wall in a tokamak 
or big molecules in the atmosphere.
Dust particles due to their large mass in comparison 
to both electrons and ions don't actively participate in DIA waves propagation 
i.e they can't resonate with the electric field oscillation.
Dust particles absorb mostly electrons and play the 
role of neutralizing background charged particles
and modify the dispersion relation.
In fact, due to the collisional process of charging, 
the dusty plasma should be considered as thermodynamically open systems\cite{ostrikov2000low,vladimirov1998evolution,vladimirov2003ion}.

In extreme cases dust particles actually absorb 
almost all electrons and a kind of dust-ion plasma emerges. 
In this case the electrons are so scarce and fully absorbed by dust particles
that they can't play any effective role 
in the dynamics of the plasma and hence in the wave propagation.
It is theoretically an interesting question
whether the ionic waves can propagate in this situation and, 
if so, what dynamics they follow.

These kind of plasma has been observed in 
both laboratory \cite{Barkan} and astrophysical plasmas.
Several plasma environments 
in the solar system, such as Titan's deep ionosphere \cite{Coates, Argen, Shebanits, Lavvas, Wellbrock},
the D-region of Earth's ionosphere\cite{Larsen, Thomas}, the E-ring plasma disk of Saturn\cite{Wahlund05, Wahlund09},
and the plume of Enceladus\cite{Morooka, Hill}.
It is also suggested through a detailed modeling t
hat strong electron depletion can be found in comet's environment such as 67P/Churyumov-Gerasimenko,
newly visited by Rosetta\cite{Vigren}. Evidences from different flybys 
(especially T70 (2004) and other 47 deep flybies -2004 until 2012-)
of Cassini have monitored the deep ionosphere of Titan.
These astronomical observations by the LP (Langmuir probe) and Cassini-RPWS (Radio and Plasma Wave Science)\cite{Lavvas} 
have shown deep decline in electron number density and it is concluded that dust/aerosol-ion plasma \cite{Shebanits} 
exists especially in the night side of the Titan. In case of Saturn's rings,
the Cassini measurements during its mission (since 2004)
have shown the existence of a decline in electron number 
density in E-ring and the plume of Enceladus \cite{Farrell}
which is an active part of the ionosphere of the this moon. 
These two environments are related since Enceladus plume 
is the main source of ionized gas for E-ring \cite{Mitchell}.

However, even in the existence of electrons,
when the electron to ion temperature ratio 
increases to a large value, then the electrons' movement 
is so fast that they can be considered 
as the neutralizing background and in this case again 
the same pattern of the ionic propagation appears
with the ions oscillating and the other constituents,
electrons and dust particles, providing the neutralizing background.

Langmuir waves along with ion acoustic and dust ion acoustic waves 
are all subject to a well-known kinetic effect named after its discoverer,
Lev Landau, as the Landau damping\cite{Landau}
which is basically damping of electric energy and converting 
it into kinetic energy without collisions involved. 
After its purely theoretical discovery, 
it has been one of the main focuses in plasma physics, 
although its physical nature and even 
mathematical proof still remains challenging 
and controversial\cite{Villani}.
However the Landau damping has been proven repeatedly 
in countless experimental and computational efforts.
As main characteristics of waves in plasmas, 
the Landau damping along with dispersion relation
are focused upon in this paper, 
and due to the pure kinetic origin of the Landau damping, 
the kinetic simulation method is employed to study these waves here.

The kinetic simulation is considered as 
the most comprehensive method to study 
many-body systems and is based on following 
the temporal evolution of distribution functions, 
which basically is the density of particles in the phase space.
The main theory governing the dynamics of distribution function 
is based on Liouville's theorem and the consistency of phase space,
which dictates the trajectories of the particles remain unique 
for each of them. By ignoring collisions, the Vlasov equation 
can be derived form the former-mentioned principle theory.
Here by following the unique trajectories of the particles 
in the phase space based on Leap-Frog scheme 
a high-precision method for the kinetic simulation is constructed.
In this method by using a randomized sampling 
in the initial step the recurrence effect is removed 
from the simulation results\cite{Jenab11E}.
Here all the three elements of dusty plasma are treated 
based on Vlasov equations and hence a fully kinetic technique is adopted.

\section{Model and numerical procedure \label{section-model}}
To follow the trajectories of the particles of different species 
in a collisionless plasma, 
the Vlasov equation as partial differential equations for each of them
\begin{multline}
\frac{\partial f_s(x,v,t)}{\partial t} + v \frac{\partial f_s(x,v,t)}{\partial x} \\ + \frac {q_s E(x,t)}{m_s} \frac{\partial f_s(x,v,t)}{\partial v} = 0 \, . \label{Vlasov}
\end{multline}
has been broken into to two ordinary 
differential equation using the method of characteristics as follows:

\begin{align*}
\frac{d x_s(t)}{d t} &=v_s\\
\frac{d v_s(t)}{d t} &= \frac {q_s E(x,t)}{m_s}
\end{align*}

in which $s=e,i,d$ stands for electrons,
ions and dust particles respectively.
$q_s$ and $m_s$ are representing charge 
and mass for each of the species. 
It is worth mentioning that since the trajectories 
are being followed in the phase space 
hence velocity and configuration space are independent from each other.
The above two equations are solved numerically based on 
a modified version of the Leap-Frog scheme\cite{Kaz}.
To follow the temporal evolution of the electric field, the Poisson equation is employed
\begin{equation}
\frac{\partial^2 \phi(x,t)}{\partial x^2} = - \frac{1}{\epsilon_0} \sum_{s = e,i,d}  \rho_s \, . \label{poisson}
\end{equation}
Finally the set of aforementioned equations are closed via density integrals over the velocity direction
\begin{equation}
\rho_s(x,t) = q_s n_{s0} \int {f_s(x,v,t)  dv}. \label{ro}
\end{equation}

Simulations are carried out in a so called 1D1V
phase space which indicates that there are one 
spatial direction $x$ and one velocity direction $v_x$.
At the initial step, the distribution function, 
which is assumed to be a Maxwellian distribution function, 
is sampled randomly by four points in each grid cell i.e. 
for each phase points which has random $x$ and $v_x$ 
a value of $f$ will be associated based on the distribution function. 
Then at each time step, the distribution function over grid points 
are constructed using the $f$ values of the phase points 
and interpolating them to the grid points.
Knowing the distribution function allows the code to calculate densities 
-by integrating the distribution function over velocity direction-
and ultimately the electric field over the spatial direction using
density integrals and the Poisson equation respectively.
In return the electric field is fed into
the Vlasov equation and the new coordinates 
($x, v_x$)
of each phase point for the next time step are calculated. 
In this method the value $f$, the phase point distribution function, 
remains untouched throughout the simulation
which avoid some numerical errors arising from constructing them.

Since the focus here is on ionic propagation, for computational reasons all the parameters are normalized to ionic parameters (the normalized quantities are marked by a acute accent) such as follows:
Space and time are scaled by $\lambda_{Di}$ and $\omega_{pi}^{-1}$ respectively, where $\omega_{pi}= \sqrt{n_{i0} e^2/(m_i \epsilon_0)}$ is the ion plasma frequency and $\lambda_{Di} = \sqrt{\epsilon_0 K_B T_i/(n_{i0} e^2)}$ is the ion Debye length. The velocity variable $v$ has been scaled by the ion thermal speed $v_{th_i} = \sqrt{K_B T_i/m_i}$, while the electric field and the electric potential have been scaled by $K_B T_i/(e \lambda_{Di})$ and $K_B T_i/e$, respectively (here $K_B$ is Boltzmann's constant). The densities of the three species are normalized by $n_{i0}$\cite{Jenab14EPJD}. The quasineutrality condition connects the three zeroth-order densities:
\begin{equation}
\sigma = 1 - \delta \, .
\end{equation}
in which $\sigma = n_{e0}/ n_{i0}$ , $\delta = Z_d n_{d0}/ n_{i0}$ are the normalized zero-order charge densities of electrons and dust particles. In order to excite an ionic propagation, the small periodic perturbation is imposed on initial distribution function i.e. Maxwellian distribution function ($f_m$) of ions.
\begin{equation}
f_i(x,v_x,t)= f_m [1+ \alpha \cos \left(\frac{2 \pi}{\lambda }x\right)].
\end{equation}

Note that quantities normalized to ionic scale are marked by acute accent $\acute{\omega}\big(=\omega/\omega_{pi}\big)$ and $\acute{k}\big(=k\lambda_{Di}\big)$, however for the sake of comparison some quantities scaled by electron parameters will be mentioned and they are marked by tilde $\widetilde{\omega}\big(=\omega/\omega_{pe}\big)$ and $\widetilde{k}\big(=k\lambda_{De}\big)$. Worth mentioning that the parameters without acute accent or tilde are parameters without any normalization.

The normalized  Bohm-Gross dispersion relation of Langmuir waves (normalized to electron parameters, $\omega$ and $k$ by $\omega_{pe}$ and $1/\lambda_{De}$ respectively
\begin{equation}
\widetilde{\omega}^2= 1+ 3 \widetilde{k}^2, \label{NBGDR}
\end{equation}
and dust ion acoustic waves (normalized to ion parameters)
\begin{equation}
\acute{\omega}^2= \acute{k}^2 \big( \frac{\theta}{1-\delta+\theta \acute{k}^2} +3   \big), \label{DIAW}
\end{equation}
are presented for the sake of comparison.

Main parameters used in this simulations are as follows:
${m_d}/{m_i} = 10^5, \quad  {m_i}/{m_e} =1836, \quad  {T_i}/{T_d}=400, \quad  Z_d=1000$. Other parameters, like $\delta$, $L= \lambda$ ($=$ the length of the simulation box), $\alpha$ and ${T_e}/{T_i}$  (hereafter $\theta$) are varied for different sets of simulations and will be addressed where appropriate. The grid applied for phase space discretization ($M_x$ points in x-direction and  $M_{vx}$ points in the x-component of the velocity) is set to $(M_x, M_{vx}) = (25, 3000)$. As a consequence, the phase space is covered by $7.5 \times 10^4$  cells, and in each of the cells, at the initial step 4 phase points are scattered randomly inside the cell. Consequently, the distribution function of each of the plasma elements is sampled by $3 \times 10^5$  points in the phase space. The number of phase points during the temporal evolution should stay unchanged since the periodic boundary condition is adopted into the code.

\section{Results and discussion \label{Section-Results}}

\begin{figure}
\centering
\includegraphics[width=7cm]{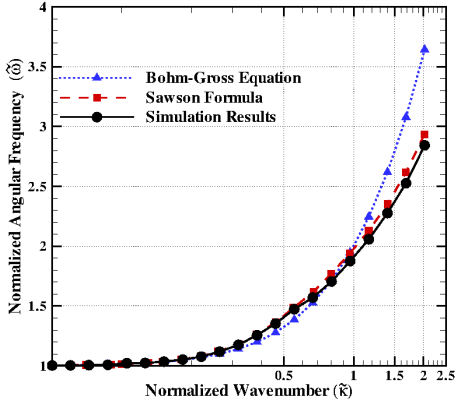}
\caption{ Comparison between analytical and simulation results -black solid curve- of dispersion relation of Langmuir waves as the benchmark. The normalization is based on electron parameters and the wavenumber axis is depicted in logarithmic scale. For the analytical results the diagrams of two equations are shown, The Bohm-Gross dispersion relation -blue dotted curve- and an empirical formula suggested by Swanson -red dashed curve.}
 \label{Fig.1}
\end{figure}

Firstly,
we employed our numerical simulation approach to study Langmuir waves 
in an ion-electron plasma in order to use the outcomes as a benchmark for our results related to the ionic perturbation.
Here a small periodic perturbation is imposed on the initial distribution function of electrons. 
The results of simulation are shown in Fig.\ref{Fig.1}, here the normalization is carried out based on electron parameters. 
In Fig.\ref{Fig.1} two analytical dispersion relations are sketched, first the Bohm-Gross dispersion relation (see Eq.\ref{NBGDR}).
The other one is an empirical formula proposed by Swanson\cite{Swason}:
\begin{equation}
\widetilde{\omega} = \big[1+\frac{1.37\widetilde{k}^2+10.4\widetilde{k}^4}{1+11.1\widetilde{k}^3}\big] \label{Swanson} 
\end{equation}
which is more accurate than the Bohm-Gross dispersion relation up to $0.2$ percent for $\widetilde{k}<0.6$.
As it is shown in the Fig.\ref{Fig.1} the simulation results match both Eq.\ref{Swanson} and Eq.\ref{NBGDR} 
for $\widetilde{k}<1.0$. However for $1.0 \leq \widetilde{k}<2.5$ a divergence occurs between the two analytical prediction, 
and the Swanson formula (Eq.\ref{Swanson}) claimed to be more precise, and expectedly 
the Bohm-Gross dispersion relation fails since it is derived for a restricted approximation of $\widetilde{k}\ll1$. 
As it is shown in Fig.\ref{Fig.1}, our simulation results agrees with the Swanson formula (Eq.\ref{Swanson}) for the regime of $1.0 \leq \widetilde{k}<2.5$ perfectly.

\begin{figure}
\centering
\includegraphics[width=7cm]{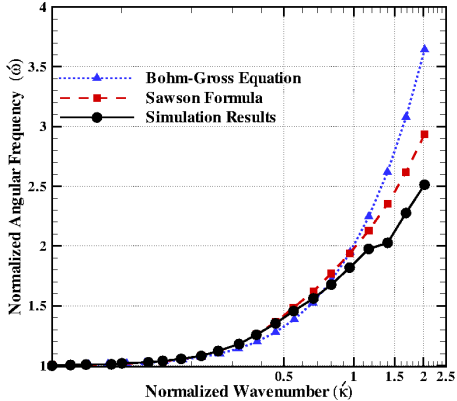}
\caption{ Normalized angular frequency is shown versus normalized wavenumber -in logarithmic scale- for $\delta = 1$ for ionic propagation. The normalization is based on ion parameters and the trends of curves can be compared to Fig. 1 which is for Langmuir waves. The simulation results -black solid curve- are sketched versus two analytical relation i.e. the Bohm-Groass equation -blue dotted curve- and empirical formula suggested by Swanson -red dashed curve.}
 \label{Fig.2}
\end{figure}

As the first step in studying the ionic propagation, we focus on the extremely electron depleted case in which electrons are mostly absorbed by dust particles so that the plasma actually consists of ions and dust particles. The results of this case $\delta =1.0$ is depicted in Fig.\ref{Fig.2} (the normalization is based on ionic parameters and the results can be compared by Fig.\ref{Fig.1} in terms of the trend but not in the scale since the scale of two figures are different, electrons/ions parameters for Fig.\ref{Fig.1}/Fig.\ref{Fig.2}). Therefore in the case of complete electron-depleted dusty plasma i.e ion and dust particle plasma, Langmuir-like ionic waves arise which show the same dispersion relation as usual Langmuir waves i.e electron Langmuir waves and attributes like the cut-off frequency ,for waves with the frequency smaller than the ion plasma frequency $\omega \ll \omega_{pi}$, clearly manifests in the dispersion relation of this wave. The transition from dust-ion-acoustic waves with no cut-off frequency to Langmuir-like ionic waves with cut-off is quite sharp, although it depends on the electron-to-ion temperature ratio as well (cf. next section). Theoretically this can be justified by setting $\delta=1$ in Eq.\ref{DIAW} which results in the Bohm-Gross dispersion relation and predicts the existence of Langmuir-like ionic waves.

\begin{figure}
\centering
\includegraphics[width=7cm]{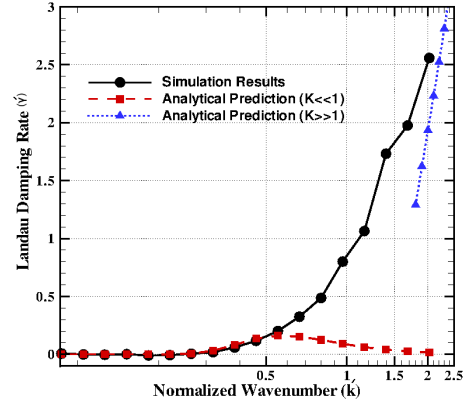}
\caption{ Normalized Landau damping rate is sketched for fully electron depleted dusty plasma $\delta=1.0$ -black solid line- and compared versus two analytical prediction one for small wavenumbers ($\acute{k}\ll 1.0$) -dashed red line- and for large wavenumbers ($\acute{k}\gg 1.0$) -dotted blue line-.}
 \label{Fig.3}
\end{figure}
The Landau damping rate as one the main features of any waves including the Langmuir-like ionic waves are presented in Fig.\ref{Fig.3}.
Simulation results are compared to two different analytical predictions\cite{Thorne}, first for $\acute{k} \ll 1$:
\begin{equation}
\acute{\gamma} =\sqrt{\frac{\pi}{8}} \frac{1}{\acute{k}^3} \exp{ \big(\frac{-1}{2\acute{k}^2}-1.5 \big)},
\end{equation}
and secondly for $\acute{k} \gg 1$:
\begin{equation}
\acute{\gamma} =\acute{k}\sqrt{2 \ln \big(\frac{1}{\sqrt{2\pi}}\acute{k}^2\big)}.
\end{equation}
The simulation results display a good agreement with both the analytical predictions in their limit of validity, for first one not beyond $\acute{k} \gg 0.5$ and for the second one for above $\acute{k} \gg 2.0$.

In the case of dusty plasma with all three elements namely electrons, ions and dust particles, setting $\delta =0.5$ as an example to start with, we focus on the case of large electron to ion temperature ratio $\theta= \frac{T_e}{T_i}\gg 1$.
\begin{figure}
\centering
\includegraphics[width=7cm]{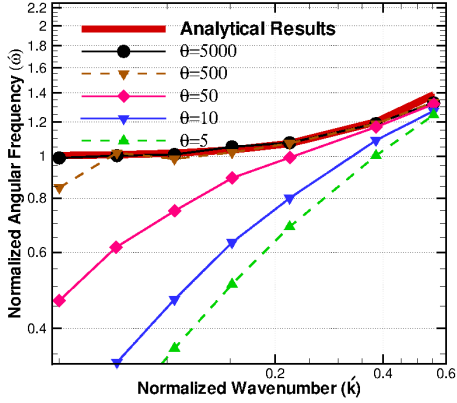}
\caption{Dispersion relation of ionic propagation is sketched for different electron to ion temperature ratio $\theta=5, 10, 50, 500$ and $5000$. As $\theta$ increases the dispersion relation converge to Langmuir waves dispersion relation which approaches zero as wavenumber drops.}
 \label{Fig.4}
\end{figure}
\begin{figure}
\centering
\includegraphics[width=7cm]{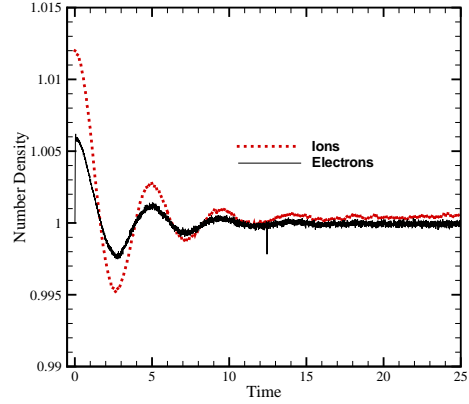}
(a)
\includegraphics[width=7cm]{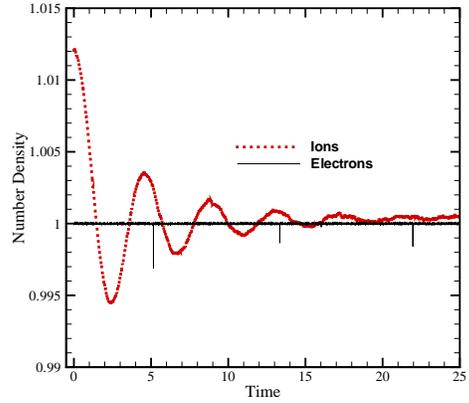}
(b)
\caption{ The number density of electrons and ions are shown for a) the case of dust ion acoustic waves with parameters $\acute{L}=11.34, \theta=5 $ and b) Langmuir-like ionic propagation with parameters $\acute{L}=11.34, \theta=5000 $ . In later case the electrons don't participate in the wave propagation and their number density don't oscillate with electric field.}
 \label{Fig.5}
 \end{figure}
Fig.\ref{Fig.4} presents the effect of the ion to electron temperature ratio on the behavior of the dispersion relation compared with the Langmuir waves dispersion relation in the case of an ionic propagation. As the temperature ratio grows the more simulation results resemble the Langmuir waves dispersion relation which indicates the existence of the Langmuir-like ionic waves discussed above.
However, in these circumstances despite the presence of electrons they don't participate in the ionic propagation unlike the ion acoustic or dust ion acoustic waves.
Here the collective movement of electrons can't follow the collective displacement of ions since electrons are moving too fast 
due to the sharp temperature ratio between them.
Fig.\ref{Fig.5} presents the temporal evolution of number densities $N(x,t)=\int f \dd v$ of electrons and ions in two different set of parameters, one for dust-ion-acoustic waves in which $\acute{L}=11.34, \theta=5$ and the other one for the Langmuir-like ionic waves with $\acute{L}=\acute{\lambda}=11.34, \theta=5000$, $\acute{L}$ and $\acute{\lambda}$ are the simulation box length and wavelength respectively. Fig.\ref{Fig.5} shows how differently electrons reacts in these two cases to the ionic propagation, in case of dust-ion acoustic waves, electrons follow the movement of ions contrary to the case of Langmuir-like ionic waves in which electrons remain passive to the wave propagation and only provide charged background like dust particles. As is the case of complete electron depleted dusty plasma, the existence of Langmuir-like ionic waves can be predicted from Eq.\ref{DIAW}. By setting $\theta \gg 1$ in Eq.\ref{DIAW} the Bohm-Gross dispersion relation Eq.\ref{NBGDR} can be achieved.

\begin{figure}
\centering
\includegraphics[width=7cm]{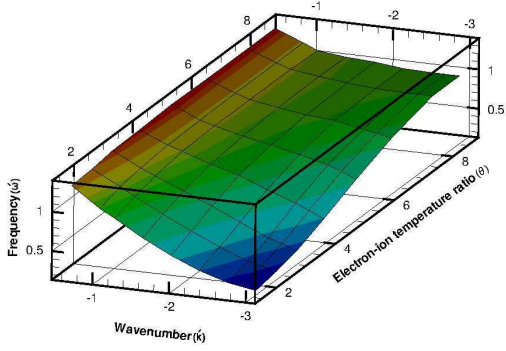}
\caption{ Dispersion relation of ionic propagation -normalized frequency over normalized wavenumber- is shown while electron to ion temperature ratio acts as the third parameter. Both normalized wavenumber and electron to ion temperature ratio $\theta$ are represented in logarithmic scale.}
 \label{Fig.6}
 \end{figure}

 \begin{figure}
\centering
\includegraphics[width=7cm]{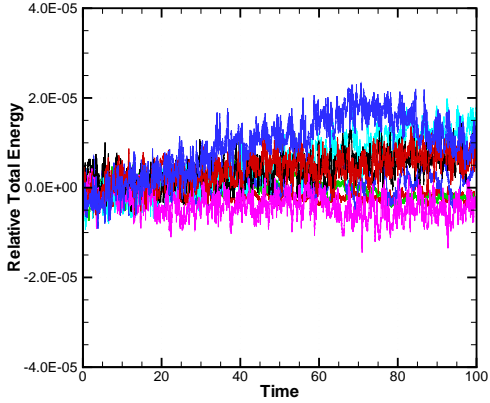}
\caption{ The conservation of total energy has been tested throughout simulation as validity criteria of simulation results. Relative total energy
$\bar{\varepsilon}(t)=\frac{\varepsilon(0)-\varepsilon(t)}{\varepsilon(0)}$
of a few simulations is sketched versus normalized time and the deviation of the initial value of the total energy is under $4\times10^{-5}$.}
 \label{Fig.7}
\end{figure}

\begin{figure}
\centering
\includegraphics[width=7cm]{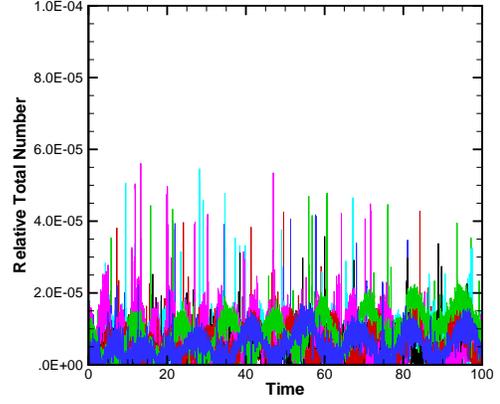}
\caption{ The conservation of total number of particles $\Pi(t)= \int {f_s(x,v,t) dv dx}$ has been considered as an essential and minimum measurement of validity of the simulation results. The relative deviation of total numbers of electrons in comparison to the initial state (t=0):
$\bar{\Pi}(t)=\frac{\Pi(0)-\Pi(t)}{\Pi(0)}$
has been presented here for of a few simulations. The deviation remains under $10^{-4}$ for whole time of the simulations.}
 \label{Fig.8}
\end{figure}

\begin{figure}
\centering
\includegraphics[width=7cm]{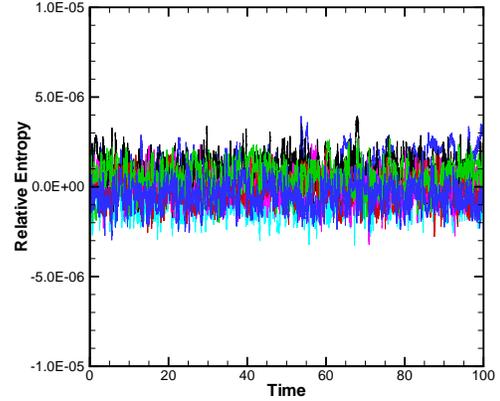}
\caption{ The conservation of entropy of electrons $S(t)= \int {f_e(x,v,t) ln f_e(x,v,t) dv dx}$ has been checked throughout the simulation as one of the yardstick of reliability of the simulation results. The relative deviation $\bar{S}(t)=\frac{S(0)-S(t)}{S(0)}$ has been shown here versus time and it remains under $10^{-5}$.}
 \label{Fig.9}
\end{figure}

Furthermore, another parameter which plays an important role is the wavenumber. Fig.\ref{Fig.6} shows the effect of the wavenumber on the dispersion relation and its connection with temperature ratio of electrons over ions. It is worth mentioning that the conservation of total energy and entropy and total numbers of particles of each species have been checked through out each simulation run and the deviation has been kept under 0.001 (cf Fig.\ref{Fig.7}, Fig.\ref{Fig.8})

\begin{figure}
\centering
\includegraphics[width=7cm]{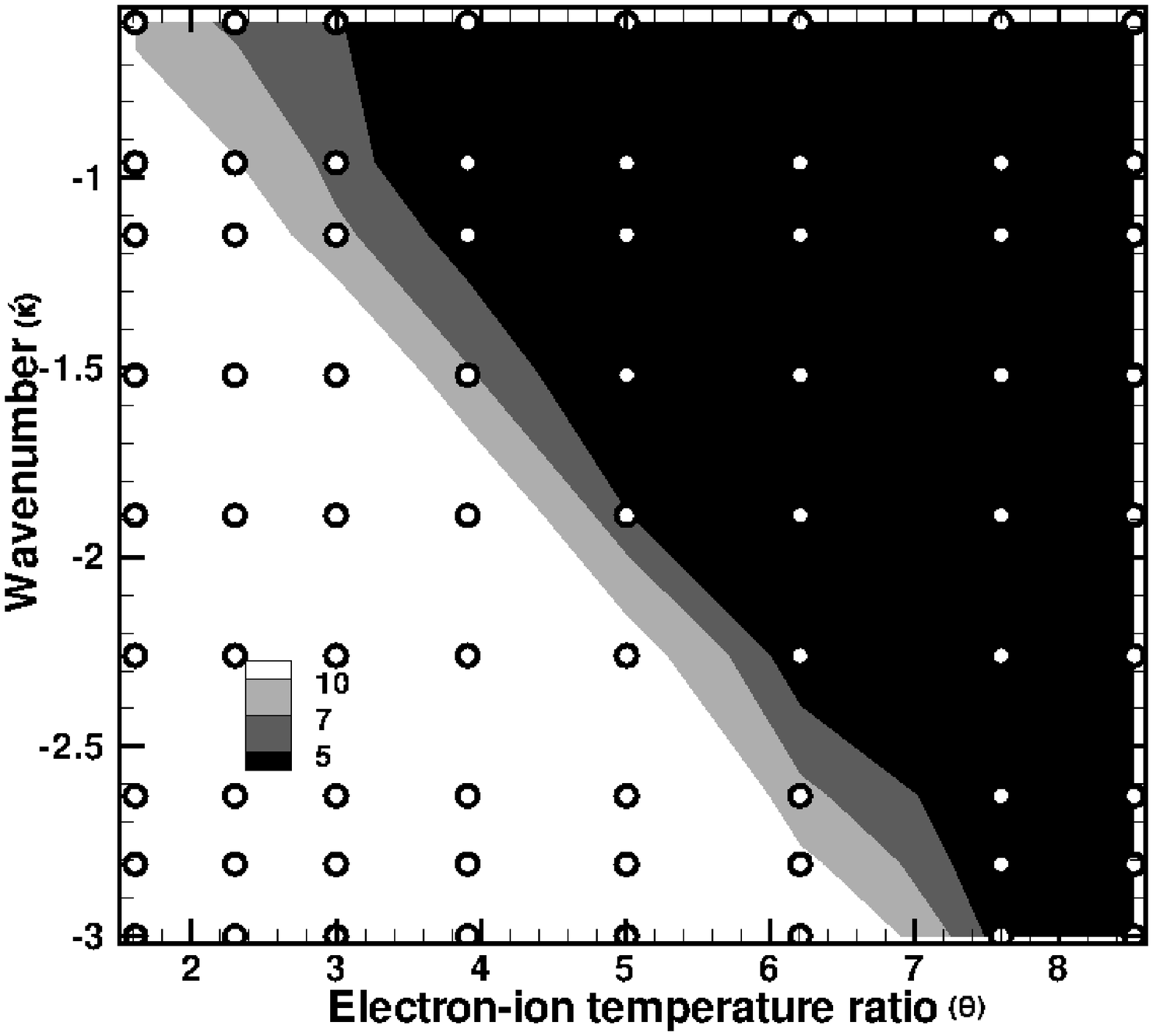}
\caption{ The relative deviation of wave frequency $\acute{\omega}$ of simulation results from analytical prediction of Langmuir waves are sketched. The area with less than 5 percent difference are shown in black. A clear transition border can be indemnified. As wavenumber and electron to ion temperature ratio grows, the dispersion relation from the simulation converges more and more into Langmuir waves dispersion relation.  }
 \label{Fig.10}
 \end{figure}
\begin{figure}
\centering
\includegraphics[width=7cm]{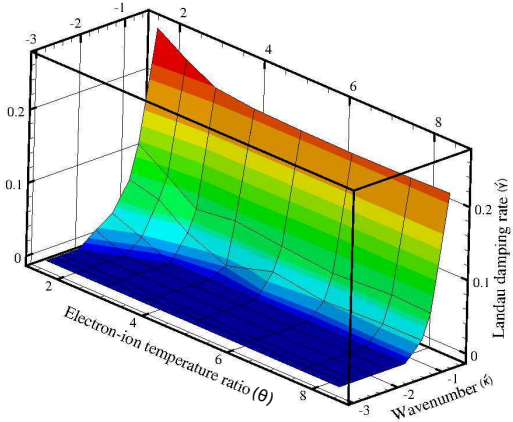}
\caption{ Normalized Landau damping rates $\acute{\gamma}$ is sketched for different values of wavenumber $\acute{k}$ and $\theta$ electron to ion temperature ratio -both in logarithmic scale - in three dimensional diagram. }
 \label{Fig.11}
\end{figure}

Fig.\ref{Fig.10} presents the relative deviation of simulation results from Langmuir waves dispersion relation
\begin{equation}
\varrho=\frac{\omega_A-\omega_S}{\omega_A}
\end{equation}
in which $\omega_A$ and $\omega_S$ presents the wave frequency of analytical predictions and simulation results respectively. In Fig.\ref{Fig.10} areas with less than 5 percent deviation are shown in black. A clear transition line can be identified in Fig.\ref{Fig.10} which depends on both wavenumber and temperature ratio of electrons over ions as it can be expected from Eq.\ref{DIAW}. This suggests what combination of two parameters namely $\delta$ and $\theta$ can produce Langmuir-like ionic waves and actually when electrons role in ionic wave propagation can be ignored. The Landau damping rate from simulation results for different combination of $\delta$ and $\theta$ is shown in Fig.\ref{Fig.11}.

However there is no sharp transition from Langmuir-like ionic waves to dust ion acoustic waves and the transition happens rather smoothly. There is no point in which two modes occur at the same time and temporal evolution of electric field gets distorted. The cut-off frequency effect doesn't exist and for each temperature there exists a threshold in wavenumber below which the dispersion relation goes below $\acute{\omega}=1$ contrary to the case of fully electron depleted  dusty plasma in which the cut-off frequency exists.

\section{Conclusions \label{Section-conclusion}}
Propagation of ionic waves in a dusty plasma has been studied for two cases i.e fully electron depleted dusty plasma and in the case of high temperature ratios of electrons over ions via fully kinetic simulation technique. The simulation results are compared to the Langmuir waves' dispersion relation and the conditions under which the ionic propagation follows Langmuir dispersion relation is shown. The existence of cut-off frequency is shown for fully electron depleted although for the second case the cut-off frequency doesn't exist. The transition from non cut-off regime of ionic waves i.e. dust ion acoustic waves to cut-off frequency regime i.e. Langmuir-like waves are studied for both cases and smooth transition is presented for the second case. The transition for the first case is sharp although it depends on temperature ratio of electrons over ions as well. The Landau damping rate as the other main feature of waves in plasmas is studied for the vast range of variables and are compared to the analytical predictions for both $\acute{k} \ll 1$ and $\acute{k} \gg 1$.

In the studied circumstances, dust particles acquire the role of heavy immobile particles delivering the quasineutrality condition by providing the negative charge background for ions which are acting as the inertia provider for the propagation by responding the electric field oscillations. For the second case the hot electrons - in comparison to ions- play the role of oppositely charged particles in background and don't actively participate in the wave propagation. It is suggested that in these two cases ionic propagation follows the dynamics similar to Langmuir waves and it can be called Langmuir-like ionic waves.

\acknowledgments
M. Jenab is thankful to Prof. I. Kourakis and Prof. M. A. Hellberg for their fruitful discussions.
This work is based upon research supported by the National Research Foundation through MWL grant no. 87616 and Department of Science and Technology.
Any opinion,findings and conclusions or recommendations expressed in this material
are those of the authors and therefore the NRF and DST do not accept any liability in regard thereto.


%

\end{document}